# Photoreflectance Study of the Fundamental Optical Properties of (Ga,Mn)As Epitaxial Films

O. Yastrubchak,[1,*] J. Żuk,[1] H. Krzyżanowska,[1,2] J. Z. Domagala,[3] T. Andrearczyk,[3] J. Sadowski,[3,4] and T. Wosinski[3]

[1]*Institute of Physics, Maria Curie-Skłodowska University, Pl. Marii Curie-Skłodowskiej 1, 20-031 Lublin, Poland*

[2]*Department of Electrical and Computer Engineering, University of Rochester, 160 Trustee Rd., Rochester, NY 14627, USA*

[3]*Institute of Physics, Polish Academy of Sciences, Al. Lotnikow 32/46, 02-668 Warszawa, Poland*

[4]*MAX-Lab, Lund University, P.O. Box 118, SE-221 00 Lund, Sweden*

**Abstract**

Fundamental optical properties of thin films of (Ga,Mn)As diluted ferromagnetic semiconductor with a low (1%) and high (6%) Mn content and of a reference GaAs film, grown by low-temperature molecular-beam epitaxy, have been investigated by photoreflectance (PR) spectroscopy. In addition, the films were subjected to complementary characterization by means of superconducting quantum interference device (SQUID) magnetometry, Raman spectroscopy, and high resolution X-ray diffractometry. Thorough full-line-shape analysis of the PR spectra, which enabled determination of the $E_0$ electronic transition in (Ga,Mn)As, revealed significant differences between the energy band structures in vicinity of the Γ point of the Brillouin zone for the two (Ga,Mn)As films. In view of the obtained experimental results the evolution of the valence band structure in (Ga,Mn)As with increasing Mn content is discussed, pointing to a merging the Mn-related impurity band with the host GaAs valence band for high Mn content.

---

[*] Corresponding author e-mail address: *yastrub@hektor.umcs.lublin.pl*

## 1. Introduction

There has been a considerable increase in the past decade in research towards developing basic materials for spin electronics. Diluted ferromagnetic semiconductors, combining semiconducting properties with magnetism, are especially promising as materials for spintronics. In this respect, (Ga,Mn)As has become a model ferromagnetic semiconductor for integrating semiconductor-based information processing and magnetic-based data storage on the same chip because of its compatibility with the established GaAs-based semiconductor technology. Homogeneous films of $Ga_{1-x}Mn_xAs$ containing up to above 10% of Mn atoms can be grown by low-temperature (200–250°C) molecular-beam epitaxy (LT-MBE).[1] When intentionally undoped the films are *p*-type, where Mn atoms, substituting Ga atoms in the GaAs crystal lattice, supply both mobile holes and magnetic moments.

On the other hand, undoped LT-MBE-grown GaAs (LT-GaAs) films are *n*-type with a rather low free electron concentration.[2] During growth an important amount (about 1%) of excess arsenic is incorporated into the GaAs matrix in the form of arsenic antisites, $As_{Ga}$, arsenic interstitials, $As_I$, and gallium vacancies, $V_{Ga}$.[2,3] These defects, mainly $As_{Ga}$ with a typical concentration of $1\times10^{20}$ cm$^{-3}$, determine the electronic structure of LT-GaAs, schematically shown in Fig. 1a.[4]

The magnetic properties of (Ga,Mn)As films arise from the Mn spin system ($S_{Mn} = 5/2$ for $Mn^{2+}$ charge state), which undergoes a ferromagnetic phase transition below the Curie temperature, $T_C$. Mn atoms substituting the Ga lattice sites in the LT-GaAs host, $Mn_{Ga}$, act as acceptors with an impurity binding energy of intermediate strength 0.11 eV.[5,6] This results in a high hole density, which is assumed to play a crucial role in the hole-mediated ordering of Mn spins.[7,8] On the other hand, Mn atoms occupying interstitial sites of the crystal lattice, $Mn_I$, act as double donors in GaAs[9] and, together with the native $As_{Ga}$ donors, partially compensate $Mn_{Ga}$ acceptors, thus resulting in effective reduction of the hole concentration in the (Ga,Mn)As films and, in turn, in decreasing their Curie temperature. With increasing the Mn content, *x*, typically both the hole concentration and $T_C$ increase in the as-grown (Ga,Mn)As films up to around $x = 0.06$, beyond which they both start to decrease, primarily due to the formation of Mn interstitials. Post-growth annealing the films at temperatures below the LT-MBE growth temperature, resulting mainly in out-diffusion of the Mn interstitials, leads to the highest $T_C$ of 185 K achieved so far for $x \approx 0.12$.[10]

$Mn_{Ga}$ acceptors are more localized than shallower, hydrogenic-like acceptors in GaAs, which results in a higher critical carrier density for the metal-insulator transition (MIT) in Mn doped GaAs, of about $1 \times 10^{20}$ cm$^{-3}$, as compared to the critical density for the shallow acceptors in GaAs which is two orders of magnitude lower.[11] Importantly, spatially homogeneous ferromagnetic ordering in (Ga,Mn)As occurs on both the insulating and metallic side of the MIT.[12]

The nature of conducting carriers mediating the ferromagnetic state in this material has not yet been unambiguously clarified. There are two alternative theories about the electronic structure of metallic (Ga,Mn)As. The first one involves persistence of the $Mn_{Ga}$-related impurity band on the metallic side of the MIT with the Fermi level, $E_f$, residing within the impurity band and mobile holes retaining the impurity band character,[13–16] as shown schematically in Fig. 1b. The second one assumes mobile holes residing in nearly unperturbed valence band of the GaAs host (Fig. 1c), which play a key role in the *p-d* Zener model of ferromagnetism in diluted magnetic semiconductors.[7,8] Elucidation of the above controversy is essential for a better understanding of carrier mediated ferromagnetism,[17] which is also important in view of potential applications of ferromagnetic semiconductors for spintronic devices. In order to resolve this question, systematic magnetic, optical, and transport experiments have been carried out on (Ga,Mn)As films by many research groups worldwide.

Most magnetic studies, as well as the trends in evolution of the $T_C$ across the phase diagram of (Ga,Mn)As, appear to be consistent with the valence band origin of the ferromagnetism in this material.[18,19] On the other hand, a number of optical results prove the impurity band character of the holes mediated in the ferromagnetic state of (Ga,Mn)As even in highly doped metallic materials.[4,14,20] Temperature-dependent dc transport data and infrared ac conductivity measurements consistently indicate the presence of an impurity band in the low-Mn-doped insulating materials and the absence of this band in (Ga,Mn)As films with the Mn content above 2%.[11] This has been interpreted in terms of the impurity-band broadening and moving closer to the valence band with increasing Mn doping and eventually merging with the valence band at higher doping. Moreover, scanning tunneling spectroscopy (STS) results have shown a significant band-gap narrowing in highly Mn-doped (Ga,Mn)As films.[21,22] Such an effect, schematically shown in Fig. 1d, may result from merging the impurity band with the host valence band into one inseparable disordered valence band[11] or from the band-gap renormalization due to many-body effect arising from the hole-hole Coulomb interaction in the (Ga,Mn)As valence band, as pointed out by Zhang and Das Sarma.[23]

In the present work, fundamental properties of the energy band structure of LT-GaAs and (Ga,Mn)As films with various Mn contents, such as interband energy transitions and electro-optic energies for light and heavy holes, were determined. Photoreflectance (PR) spectroscopy was used, as it is the most precise, nondestructive, and contactless optical characterisation technique. Owing to the derivative nature of PR it allows for accurate determination of the band-gap energies even at room temperature. The full line shape analysis of PR spectra was performed using both Airy and Aspnes third-derivative line shape (TDLS) functions. Photoreflectance studies were supported by Raman spectroscopy and high resolution X-ray diffractometry (XRD) measurements and magnetic properties of the (Ga,Mn)As films were characterized with a superconducting quantum interference device (SQUID) magnetometer.

## 2. Experimental

For our investigations two 230-nm-thick ferromagnetic (Ga,Mn)As films with Mn content of 1% and 6% have been used. The films were grown by means of the LT-MBE method at a temperature of 230°C on semi-insulating (001)-oriented GaAs substrates. Such films were chosen because of their different transport and optical properties. While (Ga,Mn)As containing 1% Mn is expected to be close to the MIT, the Mn content higher than 2% usually transfers (Ga,Mn)As films into a metallic state.[11,12,17] In addition, as a reference, we investigated a 230-nm-thick undoped GaAs film grown on GaAs by LT-MBE under the same conditions as the (Ga,Mn)As films.

Both the Mn composition and the film thickness were verified during the growth by the reflection high-energy electron diffraction (RHEED) intensity oscillations, which enabled to determine the composition and film thickness with accuracy of 0.1% and one monolayer, respectively.[24] The Mn content in (Ga,Mn)As ternary compound was calculated from the (Ga,Mn)As growth-rate increase with respect to the growth rate of a thin LT-GaAs film grown just before the (Ga,Mn)As one. Since the Mn atoms located on interstitial sites do not affect the rate of (Ga,Mn)As growth, the Mn content obtained in this way corresponds only to Mn on the Ga sites. It was determined at 1.2% and 5.9% in the two investigated (Ga,Mn)As films. However, because of a damping the RHEED oscillation amplitude during the growth, precision of Mn content determination in the films thicker than about 100 nm decreases.[24] Therefore we assume here the content of substitutional Mn atoms in the films at about 1% and 6%, respectively.

The film properties were investigated using several complementary characterization techniques. The magnetic properties and the $T_C$ values of the (Ga,Mn)As films were determined using both magnetic-field- and temperature-dependent SQUID magnetometry. Raman spectroscopy was employed to estimate the hole densities in the (Ga,Mn)As films. The micro-Raman measurements were performed using an "inVia Reflex" Raman microscope (Renishaw) at room temperature with the 514.5-nm argon ion laser line as an excitation source. The structural properties of the epitaxial films were investigated by analysing XRD data obtained at 27°C by means of high-resolution X-ray diffractometer equipped with a parabolic X-ray mirror, four-bounce Ge 220 monochromator at the incident beam and a three-bounce Ge analyzer at the diffracted beam. Misfit strain in the epitaxial films was investigated using the reciprocal lattice mapping and the rocking curve techniques for both the symmetric 004 and asymmetric 224 reflections of Cu K$\alpha_1$ radiation.

The photoreflectance experimental setup is schematically shown in Fig. 2. Room temperature PR measurements were performed using an argon ion laser (488 nm wavelength, a nominal power of 50 mW) as a pump-beam source and a 250 W halogen lamp coupled to a monochromator as a probe-beam source. The PR signal was detected by a Si photodiode. The chopping frequency of the pump beam was 70 Hz and the nominal spot size of the pump and probe beams at the sample surface was 2 mm in diameter.

### 3. Results of film characterization

Results of SQUID magnetometry applied to the (Ga,Mn)As films showed that they both exhibit an in-plane easy axis of magnetization and well-defined hysteresis loops in their magnetization vs. magnetic field dependence, as shown in the inset in Fig. 3. The films displayed a $T_C$ value of 40 K and 60 K for a Mn content of 1% and 6%, respectively, as obtained from temperature-dependent magnetization results shown in Fig. 3. This range of Curie temperature is typical for unannealed, relatively thick (Ga,Mn)As films.[4] The shape of temperature-dependent magnetization curves differs from the universal curve for conventional ferromagnets but is typical for (Ga,Mn)As films and is related to temperature-dependent magnetic anisotropy of the films (cf. Refs. 12 and 19).

The micro-Raman spectra obtained for the LT-GaAs and (Ga,Mn)As films are presented in Fig. 4. Quantitative analysis of Raman spectra can provide important information about the free-carrier density. Seong et al.[25] proposed a powerful procedure, which enables an accurate determination of the carrier density without applying large magnetic fields, as is required in the Hall-effect measurements for ferromagnetic materials. In (Ga,Mn)As films, characterized by a high density of free holes of about $10^{20}$ cm$^{-3}$, the interaction between the hole plasmon and the LO phonon leads to the formation of coupled plasmon–LO phonon (CPLP) mode.[26] In addition, it results in a broadening and a shift of the Raman line from the LO-phonon position to the TO-phonon position depending on the hole density.[27] By performing a full line-shape analysis of our micro-Raman spectra obtained for the (Ga,Mn)As films, shown in Fig. 4, and comparing them with the published data,[26,27] we were able to estimate a hole concentration of $0.9\times10^{20}$ cm$^{-3}$ and $1.4\times10^{20}$ cm$^{-3}$ in the films with the Mn content of 1% and 6%, respectively. These results suggest that the first (Ga,Mn)As film demonstrates properties of an insulator-like material and the second one - those of a metallic-like material.[11]

Our results of high-resolution XRD measurements revealed that both the LT-GaAs and (Ga,Mn)As films, grown on a GaAs substrate under compressive misfit stress, were fully strained to the (001) GaAs substrate. The clear X-ray interference fringes, observed for the 004 Bragg reflections shown in Fig. 5, indicate the high structural perfection of the films. The film thicknesses calculated from the angular spacing of the fringes correspond very well to their thicknesses determined from the growth parameters. Angular positions of the peaks corresponding to the LT-GaAs and (Ga,Mn)As films were used to calculate the perpendicular lattice parameters, $c$, the relaxed lattice parameters, $a_{rel}$, (assuming the (Ga,Mn)As elasticity constants to be the same as for GaAs), the lattice mismatch with respect to the GaAs substrate, and the vertical strain. The results are listed in Table 1. The lattice unit of the films changes with increasing lattice mismatch from the zinc-blende cubic structure to the tetragonal structure with the perpendicular lattice parameter larger than the lateral one, equal to the GaAs lattice parameter. The observed broadening of the diffraction peak corresponding to the $Ga_{94}Mn_{06}As$ film (Fig. 5) indicates a spread of its perpendicular lattice parameter, which may result from slight lateral inhomogeneity of the Mn content along the sample surface. In fact, additional XRD measurements, performed with a very narrow X-ray beam, confirmed a lateral inhomogeneity of the Mn content in the film of up to 0.4%. Other possible reasons of the peak broadening, as e.g. vertical gradient of the film composition, were excluded in view of our results of the reciprocal lattice mapping performed with a narrow X-ray beam for the 004 reflection.

### 4. Photoreflectance results and discussion

The photoreflectance study presented in this paper enabled determination of the evolution of the fundamental energy gap in (Ga,Mn)As films with increasing Mn content. The PR spectra measured in the photon-energy range from 1.35 to 1.70 eV for both the LT-GaAs and (Ga,Mn)As epitaxial films are shown in Fig. 6. As the PR signal is associated with the electric field caused by the separation of photogenerated charge carriers, its intensity significantly decreases in highly doped semiconductors because of efficient screening the electric field by free carriers. Accordingly, intensity of the measured PR signal strongly decreased with increasing Mn content in the investigated films. This intensity for the $Ga_{94}Mn_{06}As$ sample was smaller by about one order of magnitude than that for the LT-GaAs sample. The spectra presented in Fig. 6 have been normalized to the same intensity. All the experimental spectra reveal a rich, modulated structure containing three main features: electric-field-induced Franz-Keldysh oscillations (FKO) at energies above the fundamental absorption edge, a peak at around the GaAs energy gap, and a below-band-gap feature.

The thicknesses of our epitaxial films are larger than the penetration depth of 488-nm argon-ion-laser line in crystalline GaAs, estimated to about 100 nm.[28] Nevertheless, we were able to detect a PR-signal contribution coming from the film-substrate interface regions. This contribution results from the transport of laser-injected carriers to the interface and formation of space-charge volume, which extends in GaAs to 1 μm in depth. Sydor et al.[29] performed detailed study of the nature of below-band-gap feature in their PR spectra from GaAs thin films MBE-grown on GaAs substrates and related it to impurity effects at the GaAs/GaAs interface. According to their interpretation, the modulation mechanism of this feature results from the thermal excitation of impurities or traps at the interface and their momentary refilling by the laser-injected carriers. In line with this interpretation, the relative amplitude of the below-band-gap feature in our PR results shown in Fig. 6 increases with the Mn content in our epitaxial films.

Detailed numerical analysis of experimental spectra, performed by using Airy and Aspnes TDLS functions,[30] allowed us to determine fundamental parameters, like energies of interband transitions, electro-optic energies for light and heavy holes and the ratio of the light- and heavy-hole interband reduced masses. Because of the excitonic nature of the room-temperature PR signal, the measured energies for interband transitions in GaAs are smaller than the nominal band-gap energies by approximately the excitonic binding energy. Thus, the energies obtained from a TDLS analysis are called critical-point energies, $E_{CP}$, rather than the band-gap energies. The critical-point energy corresponding to the fundamental band-gap transition at the Γ point of the Brillouin zone is denoted by $E_0$. In general, the normalized change in reflection measured in PR, $\Delta R/R$, is defined as:[31]

$$\Delta R / R = \alpha \Delta \varepsilon_1 + \beta \Delta \varepsilon_2, \qquad (1)$$

where $\alpha$ and $\beta$ are the Seraphin coefficients,[31] and $\Delta\varepsilon_1$ and $\Delta\varepsilon_2$ are the pump-laser-induced changes in real and imaginary dielectric functions, respectively. In GaAs, near the $E_0$ critical point, the values of $\beta$ coefficient are very close to zero.[31] Consequently, the second term in eq. (1) was neglected during the fitting procedures. Equation (1) then reduces to: $\Delta R/R = \alpha \Delta \varepsilon_1$. The change in the real part of the dielectric function is expressed by:[32] $\Delta \varepsilon_1 = B \theta^{1/2} \operatorname{Im}\left[\dfrac{H(z)}{(\hbar\omega - i\gamma_0)^2}\right]$, where:

$$H(z) = 2\pi[\exp((-\pi/e)i)[A_i'(z)A_i'(w) + wA_i(z)A_i(w)]] - \left(\dfrac{-\eta + (\eta^2 + \gamma^2)^{1/2}}{2}\right)^{1/2} + i\left(\dfrac{\eta + (\eta^2 + \gamma^2)^{1/2}}{2}\right)^{1/2}$$

, $B$ is a constant related to the polarization and transition strength, the electro-optic energy $\hbar\theta = \left(\dfrac{e^2\hbar^2 F^2}{2\mu}\right)^{1/3}$, where $F$ is the electric field and $\mu$ is the interband reduced effective mass, $\hbar\omega$ is the photon energy of the probe beam, $\gamma_0$ is a broadening parameter, $\gamma$ is expressed by $\gamma_0/\hbar\theta$, and $A_i$ and $A_i'$ are Airy functions and their derivatives, respectively. The arguments $z$ and $w$ of the Airy functions are given by the following formulas: $z = \eta + i\gamma = \dfrac{(E_{CP} - \hbar\omega)}{\hbar\theta} + i\dfrac{\gamma_0}{\hbar\theta}$ and $w = z\exp((-2\pi/3)i)$.

Additionally, the three-dimensional TDLS, proposed by Aspnes,[30] was utilized to fit the below-band-gap feature:

$$\Delta R/R = \operatorname{Re}[C\exp(i\vartheta)(\hbar\omega - E_{CP} + i\gamma)^{-n}], \qquad (2)$$

where $C$ is an amplitude parameter, $\vartheta$ is the phase parameter, and index $n = 2.5$ represents the three-dimensional band-to-band transition. The Franz-Keldysh oscillations were fitted using the two Airy line-shape functions to take into account both light-hole, $lh$, and heavy-hole, $hh$, valence bands.[32]

In consequence, the relative changes in reflection coefficient were estimated using the following terms during the fitting procedure:

$$\Delta R/R = \alpha \Delta \varepsilon_1(Ga_{1-x}Mn_xAs, lh) + \alpha \Delta \varepsilon_1(Ga_{1-x}Mn_xAs, hh)$$
$$+ TDLS\ (Ga_{1-x}Mn_xAs/GaAs\ interface).$$

Calculations were done using the Levenberg-Marquardt algorithm with the Matlab 8.1 computer code. The functional form of the Seraphin coefficient $\alpha$ was applied. The fits are represented by solid lines in Fig. 6. Table 2 summarizes the best fit parameters for all the heterostructures studied. The $E_0$ value obtained for the LT-GaAs film is in good agreement with earlier results obtained for LT-GaAs (cf. Ref. 33), which confirms a correctness of the fitting procedure used here.

Thorough analysis of the results shows the slight blue shift, of 4 meV of the $E_0$ transition energy in the $Ga_{0.99}Mn_{0.01}As$ film, relative to the reference LT-GaAs film. In contrast, a substantial red shift of 40 meV, of $E_0$ was revealed for the $Ga_{0.94}Mn_{0.06}As$ film as seen in Table 2 where $E_0 = 1.386$ eV. Such a sequence of the obtained $E_0$ values has also been confirmed by using a simplified approach consisting in an analysis of the energy positions of FKO extrema in the PR spectra. This approach neglects the light-hole contribution to PR spectra and takes the asymptotic expression of the Airy function.[34] The band-gap energies are then obtained from the intersection with ordinate of the linear dependence of the energies of FKO extrema vs. their "effective index" defined as $F_j = [3\pi(j-1/2)/2]^{2/3}$, where $j$ is the extremum number.[29]

The significant reduction of the energy gap in highly Mn-doped (Ga,Mn)As film is in qualitative agreement with the results of STS,[21,22] which suggested even smaller band gap of 1.23 eV in (Ga,Mn)As film with 3.2% Mn content, as estimated from conductance spectra measured in a scanning tunneling microscope.[21] The large difference between that result and our experimental value for $E_0$ may result from different measurement techniques.

On the other hand, the blue shift of the $E_0$ energy in the $Ga_{0.99}Mn_{0.01}As$ film is in agreement with recent PR investigations performed by Alberi et al.[16] for epitaxial films of (Ga,Mn,Be)As quaternary alloy with the Mn content in the range from 2.5% to 3.8%. The above authors reported on an increase of the $E_0$ energy with increasing Mn content in the investigated range and interpreted their results in a model of an anticrossing-induced formation of a Mn impurity band. As a result of codoping the films with Be shallow acceptors they displayed rather high free hole concentration of about $5 \times 10^{20}$ cm$^{-3}$ independent of Mn content. However, "codoping of (Ga,Mn)As with Be acceptors creates a huge increase of Mn interstitials, thus destroying ferromagnetism" as was concluded in a previous paper of similar group of authors.[35] This property of the (Ga,Mn,Be)As films and/or the limited range of Mn content in the films investigated by Alberi et al.[16] may be the reason for a discrepancy between their results and our results for the (Ga,Mn)As film with 6% Mn content.

Our PR results evidence a difference in the electronic band structures of (Ga,Mn)As of insulator- ($p = 0.9 \times 10^{20}$ cm$^{-3}$) and metallic-like ($p = 1.4 \times 10^{20}$ cm$^{-3}$) types of materials. All the parameters of the band structure of the 1% Mn-doped (Ga,Mn)As film in vicinity of the Γ point, revealed from the PR results and listed in Table 2, are similar to those of the LT-GaAs film. The slight blue shift of $E_0$ transition in the Ga$_{0.99}$Mn$_{0.01}$As film indicates the Fermi level position below the top of the GaAs valence band, which is consistent with the situation presented in Fig. 1c, where the $E_0$ transition occurs from the Fermi level to the conduction band. Then, the small increase in $E_0$ may result from the Moss-Burstein shift of the absorption edge.[36]

On the other hand, the significant red shift of the $E_0$ transition in the (Ga,Mn)As film with 6% Mn content is consistent with the band structure presented in Fig. 1d, where the Mn-related impurity band is merged with the GaAs valence band, forming a disordered valence band extending within the band gap.[11] In this case the $E_0$ transition occurs from the Fermi level in the disordered valence band to the conduction band. The magnitude of the red-shift of the $E_0$ transition results from the interplay between the band-gap narrowing and the Moss-Burstein shift in highly Mn-doped (Ga,Mn)As. Owing to the disordered valence band, all the parameters obtained from the PR spectra for the Ga$_{0.94}$Mn$_{0.06}$As film, listed in Table 2, differ markedly from those of the LT-GaAs and Ga$_{0.99}$Mn$_{0.01}$As films. Moreover, a lack of splitting of the PR spectra into light- and heavy-hole features in the spectral area near the $E_0$ transition, even in the $Ga_{0.94}Mn_{0.06}As$ film with a vertical strain as high as $4.2\times10^{-3}$ (Table 1), may be explained by the disordered character of valence band in this case.

## 5. Conclusions

In this work we have employed complementary characterization techniques, such as photoreflectance spectroscopy, Raman spectroscopy, high resolution X-ray diffractometry, and SQUID magnetometery, for revealing the fundamental properties of (Ga,Mn)As epitaxial films with different Mn content. Our PR spectroscopy measurements, supported with the full line shape analysis of PR spectra, enabled determination of the $E_0$ electronic transition in (Ga,Mn)As and its dependence on the Mn content. In (Ga,Mn)As with 1% Mn content and hole density close to that of the MIT, the $E_0$ energy was slightly blue shifted with respect to that in reference LT-GaAs, which was interpreted as a result of the Moss-Burstein shift of the absorption edge due to the Fermi level location below the top of GaAs valence band. On the other hand, a substantial red shift, of 40 meV, of the $E_0$ energy was revealed in (Ga,Mn)As with a high (6%) Mn content and a hole density corresponding to metallic side of the MIT. This result, together with the determined other parameters of the intrreband electro-optic transitions near the center of the Brillouin zone, which were significantly different from those in reference LT-GaAs, was interpreted in terms of a disordered valence band, extended within the band-gap, formed in highly Mn-doped (Ga,Mn)As as a result of merging the Mn-related impurity band with the host GaAs valence band.

## Acknowledgments

O. Y. acknowledges financial support from the Polish Ministry of Science and Higher Education (MSHE) under Grant POL-POSTDOC III, N PBZ/MNiSW/07/2006/33. H. K. wishes to thank MSHE for financial support under Grant 224/MOB/2008/0. This work was also supported by MSHE Project No. N N202 129339. The MBE project at MAX-Lab is supported by the Swedish Research Council (VR)

# References

1. F. Matsukura, H. Ohno, A. Shen, and Y. Sugawara, Phys. Rev. B **57**, R2037 (1998).
2. D.C. Look, D.C. Walters, M.O. Manasreh, J.R. Sizelove, C.E. Stutz, and K.R. Evans, Phys. Rev. B **42**, 3578 (1990).
3. M. Kaminska, Z. Liliental-Weber, E.R. Weber, T. George, J.B. Kortright, F.W. Smith, B.-Y. Tsaur, and A.R. Calawa, Appl. Phys. Lett. **54**, 1881 (1989).
4. E.J. Singley, K.S. Burch, R. Kawakami, J. Stephens, D.D. Awschalom, and D.N. Basov, Phys. Rev. B **68**, 165204 (2003).
5. R.A. Chapman and W.G. Hutchinson, Phys. Rev. Lett. **18**, 443 (1967).
6. A.M. Yakunin, A.Yu. Silov, P.M. Koenraad, J.H. Wolter, W. Van Roy, J. De Boeck, J.-M. Tang, and M.E. Flatté, Phys. Rev. Lett. **92**, 216806 (2004).
7. T. Dietl, H. Ohno, and F. Matsukura, Phys. Rev. B **63**, 195205 (2001).
8. T. Jungwirth, J Sinova, J. Mašek J. Kučera, and A.H. MacDonald, Rev. Mod. Phys. **78**, 809 (2006).
9. K.M. Yu, W. Walukiewicz, T. Wojtowicz, I. Kuryliszyn, X. Liu, Y. Sasaki, and J.K. Furdyna, Phys. Rev. B **65**, 201303(R) (2002).
10. M. Wang, R.P. Campion, A.W. Rushforth, K.W. Edmonds, C.T. Foxon, and B.L. Gallagher, Appl. Phys. Lett. **93**, 132103 (2008).
11. T. Jungwirth, J. Sinova, A.H. MacDonald, B.L. Gallagher, V. Novák, K.W. Edmonds, A.W. Rushforth, R.P. Campion, C.T. Foxon, L. Eaves, E. Olejník, J. Mašek, S.-R. Eric Yang, J. Wunderlich, C. Gould, L.W. Molenkamp, T. Dietl, and H. Ohno, Phys. Rev. B **76**, 125206 (2007).
12. S.R. Dunsiger, J.P. Carlo, T. Goko, G. Nieuwenhuys, T. Prokscha, A. Suter, E. Morenzoni, D. Chiba, Y. Nishitani, T. Tanikawa, F. Matsukura, H. Ohno, J. Ohe, S. Maekawa, and Y.J. Uemura, Nature Mat. **9**, 299 (2010).
13. M. Berciu, and R.N. Bhatt, Phys. Rev. B **69**, 045202 (2004).
14. K.S. Burch, D.B. Shrekenhamer, E.J. Singley, J. Stephens, B.L. Sheu, R.K. Kawakami, P. Schiffer, N. Samarth, D.D. Awschalom, and D.N. Basov, Phys. Rev. Lett. **97**, 087208 (2006).
15. B.L. Sheu, R.C. Myers, J.-M. Tang, N. Samarth, D.D. Awschalom, P. Schiffer, and M.E. Flatté, Phys. Rev. Lett. **99**, 227205 (2007).
16. K. Alberi, K.M. Yu, P.R. Stone, O.D. Dubon, W. Walukiewicz, T. Wojtowicz, X. Liu, and J.K. Furdyna, Phys. Rev. B **78**, 075201 (2008).
17. A. Richardella, P. Roushan, S. Mack, B. Zhou, D.A. Huse, D.D. Awschalom, and A. Yazdani, Science **327**, 665 (2010).

18. T. Jungwirth, K.Y. Wang, J. Mašek, K.W. Edmonds, J. König, J. Sinova, M. Polini, N.A. Goncharuk, A.H. MacDonald, M. Sawicki, A.W. Rushforth, R.P. Campion, L.X. Zhao, C.T. Foxon, and B.L. Gallagher, Phys. Rev. B **72**, 165204 (2005).
19. M. Sawicki, D. Chiba, A. Korbecka, Y. Nishitani, J.A. Majewski, F. Matsukura, T. Dietl, and H. Ohno, Nature Phys. **6**, 22 (2010).
20. V.F. Sapega, M. Ramsteiner, O. Brandt, L. Däweritz, and K.H. Ploog, Phys. Rev. B, **73**, 235208 (2006).
21. T. Tsuruoka, N. Tachikawa, S. Ushioda, F. Matsukura, K. Takamura, and H. Ohno, Appl. Phys. Lett. **81**, 2800 (2002).
22. G. Mahieu, P. Condette, B. Grandidier, J.P. Nys, G. Allan, D. Stiévenard, Ph. Ebert, H. Shimizu, and M. Tanaka, Appl. Phys. Lett. **82**, 712 (2003).
23. Y. Zhang and S. Das Sarma, Phys. Rev. B **72**, 125303 (2005).
24. J. Sadowski J. Z. Domagała, J. Bąk-Misiuk, S. Koleśnik, M. Sawicki, K. Świątek, J. Kanski, L. Ilver, and V. Ström, J. Vac. Sci. Technol. B **18**, 1697 (2000).
25. M.J. Seong, S.H. Chun, H.M. Cheong, N. Samarth, and A. Mascarenhas, Phys. Rev. B **66**, 033202 (2002).
26. G. Irmer, M. Wenzel, and J. Monecke, Phys. Rev. B **56,** 9524 (1997).
27. W. Limmer, M. Glunk, W. Schoch, A. Köder, R. Kling, R. Sauer, and A. Waag, Physica E **13,** 589 (2002).
28. O. Madelung (Ed.), *Data in Science and Technology, Semiconductors, Group IV Elements and III-V Compounds*, Springer-Verlag, Berlin, 1991.
29. M. Sydor, J. Angelo, J.J. Wilson, W.C. Mitchel, and M.Y. Yen, Phys. Rev. B **40**, 8473 (1989).
30. D.E. Aspnes, Surf. Sci. **37**, 418 (1973).
31. B.O. Seraphin and N. Bottka, Phys. Rev. **145**, 628 (1966).
32. J.P. Estrera, W.M. Duncan, and R. Glosser, Phys. Rev B **49,** 7281 (1994).
33. S. Sinha, B.M. Arora, and S. Subramanian, J. Appl. Phys. **79**, 427 (1996).
34. D.E. Aspnes and A.A. Studna, Phys. Rev. B **7**, 4605 (1973).
35. K.M. Yu, W. Walukiewicz, T. Wojtowicz, W.L. Lim, X. Liu, U. Bindley, M. Dobrowolska, and J.K. Furdyna, Phys. Rev. B **68**, 041308(R) (2003).
36. J. Szczytko, W. Mac, A. Twardowski, F. Matsukura, and H. Ohno, Phys.Rew. B **59**, 12935 (1999).

Table 1.

Lattice parameters, lattice mismatch (defined as $(a_{rel} - a_{sub})/a_{sub}$, where $a_{sub}$ means the lattice constant of GaAs substrate), and vertical strain (defined as $(c - a_{rel})/a_{rel}$) for the investigated films calculated from the results of XRD measurements performed at 27°C.

| sample | 2θ (±0.003) | $c$ [Å] (±0.00008) | $a_{rel}$ [Å] | lattice mismatch [×$10^4$] | vertical strain [×$10^4$] |
|---|---|---|---|---|---|
| GaAs | | 5.65349 | | | |
| LT-GaAs | 65.9915 | 5.65789 | 5.65563 | 3.80 | 3.99 |
| $Ga_{0.99}Mn_{0.01}As$ | 65.897 | 5.66509 | 5.65914 | 10.00 | 10.51 |
| $Ga_{0.94}Mn_{0.06}As$ | 65.4426 | 5.70001 | 5.67616 | 40.11 | 42.02 |

Table 2.

Values of energy transitions, electro-optic energies for light and heavy holes, the ratio of the light- and heavy-hole interband reduced masses, and broadening parameters for the investigated heterostructures obtained from fitting procedures of the PR spectra.

| parameter | reference sample | | $Ga_{0.99}Mn_{0.01}As$ | | $Ga_{0.94}Mn_{0.06}As$ | |
| | LT-GaAs/ GaAs interface | LT-GaAs film | (Ga,Mn)As/ GaAs interface | (Ga,Mn)As film | (Ga,Mn)As/ GaAs interface | (Ga,Mn)As film |
|---|---|---|---|---|---|---|
| $E_{CP}$ or $E_0$ (eV) | 1.417 | 1.426 | 1.421 | 1.430 | 1.420 | 1.386 |
| $\hbar\theta_{lh}$ (meV) | - | 48.29 | - | 50.13 | - | 57.53 |
| $\hbar\theta_{hh}$ (meV) | - | 37.25 | - | 38.27 | - | 34.37 |
| $\mu_{lh}/\mu_{hh}$ | - | 0.46 | - | 0.44 | - | 0.21 |
| $\gamma$ (eV) | 0.011 | 0.031 | 0.018 | 0.022 | 0.030 | 0.058 |

**Figure captions**

Fig. 1. Schematic energy band diagram for LT-GaAs (a) and possibilities of its evolution for (Ga,Mn)As with increasing Mn content (b, c, and d). Splitting of the bands in the ferromagnetic state is omitted for simplicity. Arrows indicate electronic transitions from the valence band to the conduction band ($E_0$). Impurity-band regime for low-Mn-doped (Ga,Mn)As is presented in (b): a narrow impurity band is formed at an energy of the $Mn_{Ga}$ acceptor level, separated from the valence band by an energy gap of the magnitude close to the impurity binding energy, with the Fermi level residing inside this band (assuming some compensation). An alternative scenario is shown (c), where holes are doped into the valence band from $Mn_{Ga}$ acceptor levels and the absorption edge shifts from the center of the Brillouin zone to the Fermi-wave vector. The disordered-valence-band regime for high-Mn-doped (Ga,Mn)As is presented in (d), where the impurity band and the host valence band merge into one inseparable band, whose tail may still contain localized states (shaded grey area) depending on the free carrier concentration and disorder.

Fig. 2. Scheme of the photoreflectance experimental arrangement.

Fig. 3. SQUID magnetization along the in-plane [110] crystallographic direction vs. temperature for the (Ga,Mn)As films after subtraction of diamagnetic contribution from the GaAs substrate. Inset: magnetization hysteresis loops measured at a temperature of 5 K.

Fig. 4. Raman spectra recorded at room temperature in backscattering configuration from the (001) surfaces of the LT-GaAs reference film and the two (Ga,Mn)As films. The spectra have been vertically offset for clarity. The dashed lines indicate the positions of the Raman LO- and TO-phonon lines for the LT-GaAs reference film.

Fig. 5. X-ray diffraction results: 2θ/ω scans for (004) Bragg reflections for 230-nm LT-GaAs, $Ga_{0.99}Mn_{0.01}As$, and $Ga_{0.94}Mn_{0.06}As$ thin films grown on (001) semi-insulating GaAs substrate. The curves have been vertically offset for clarity. The narrow line corresponds to reflection from the GaAs substrate and the broader peaks at lower angles are reflections from the epitaxial films. With increasing the Mn content the (Ga,Mn)As diffraction peaks shift to smaller angles with respect to that of the LT-GaAs reference film.

Fig. 6. Sequence of the photoreflectance spectra for the LT-GaAs, $Ga_{0.99}Mn_{0.01}As$, and $Ga_{0.94}Mn_{0.06}As$ films epitaxially grown on GaAs substrate (dots). Solid lines represent fits to the experimental data. The spectra have been normalized to the same intensity and vertically offset for clarity. Vertical arrows indicate the $E_0$ transitions in the epitaxial films determined from fitting procedures of the whole PR spectra.

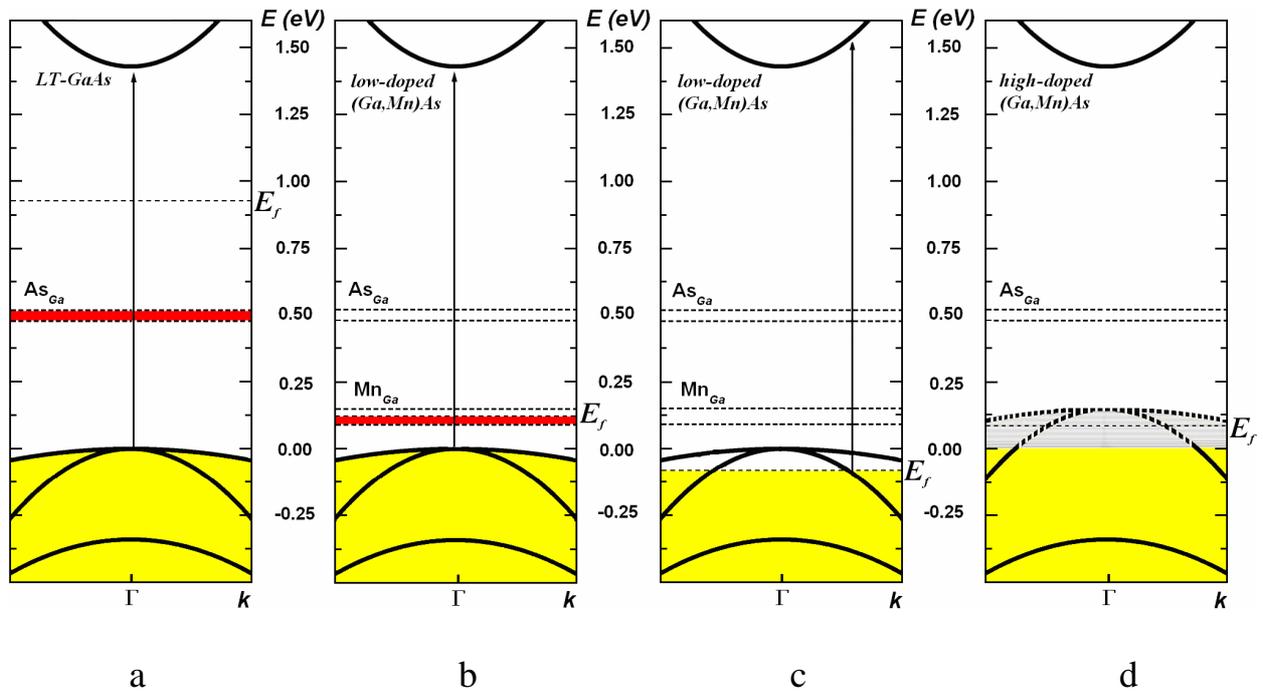

Fig. 1

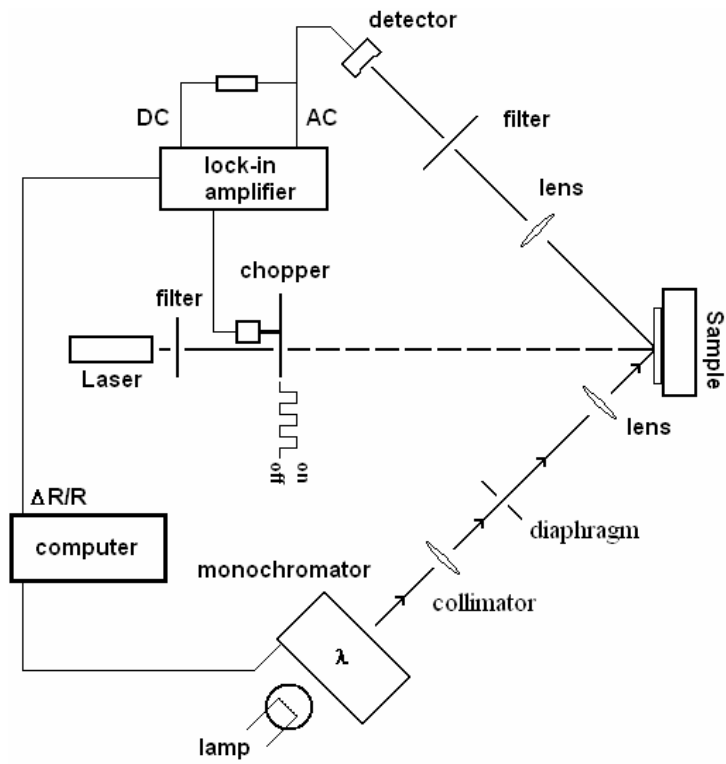

Fig. 2

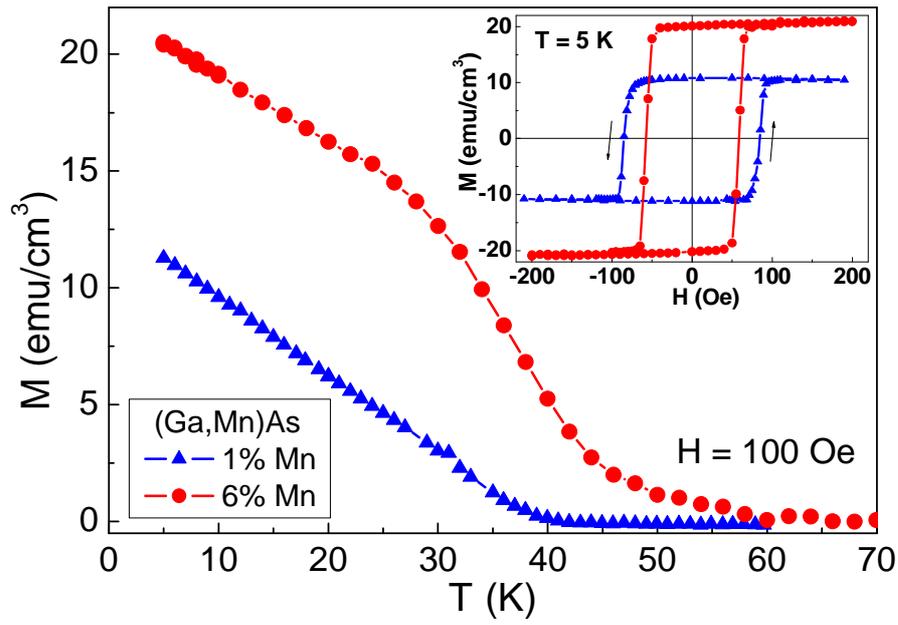

Fig. 3

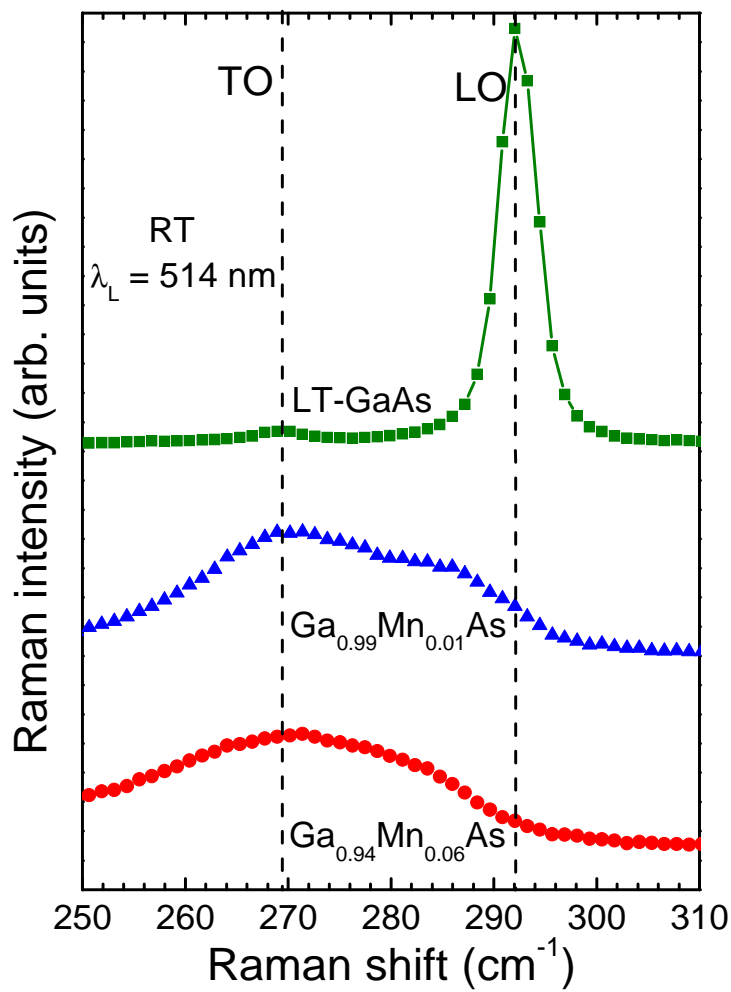

Fig. 4

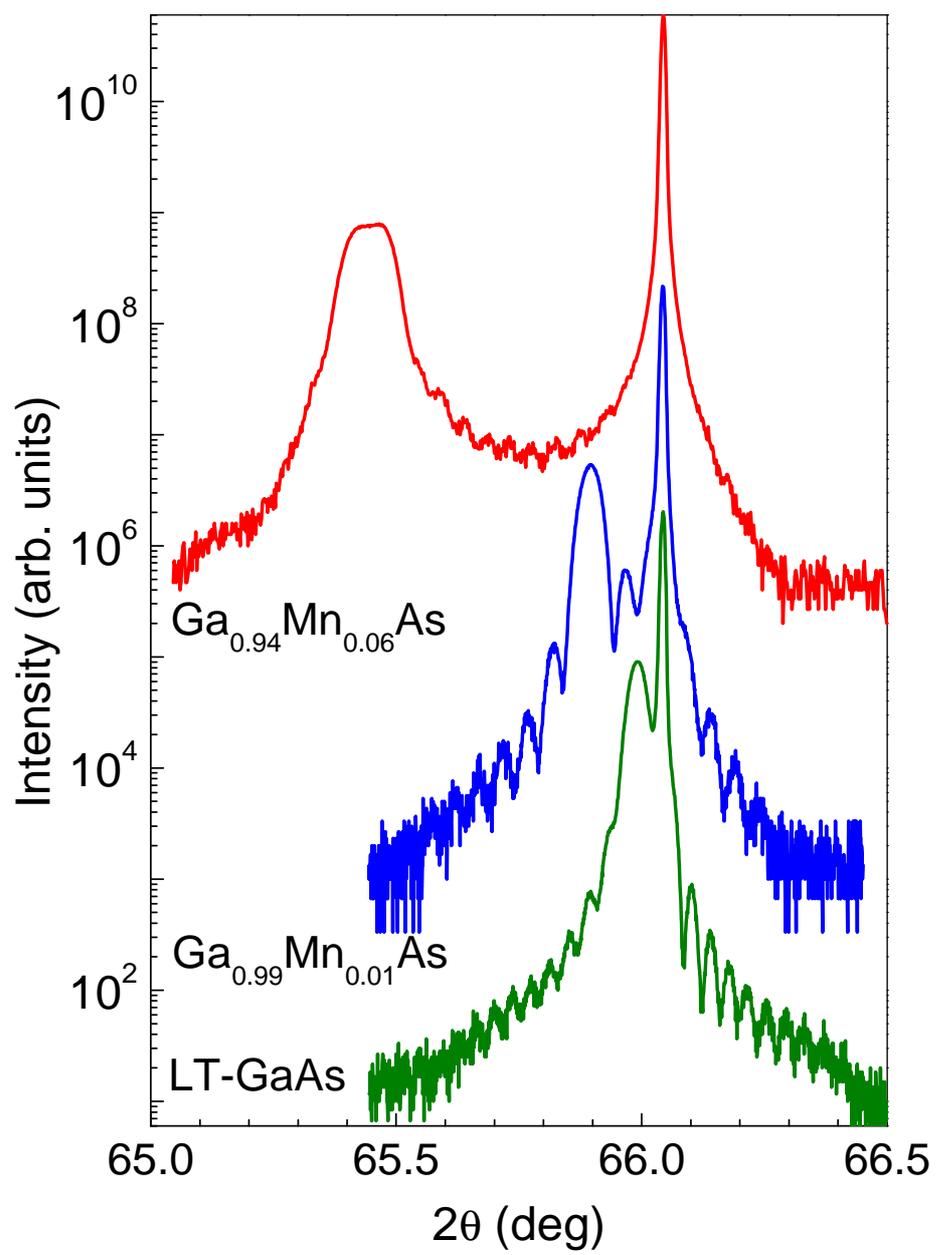

Fig. 5

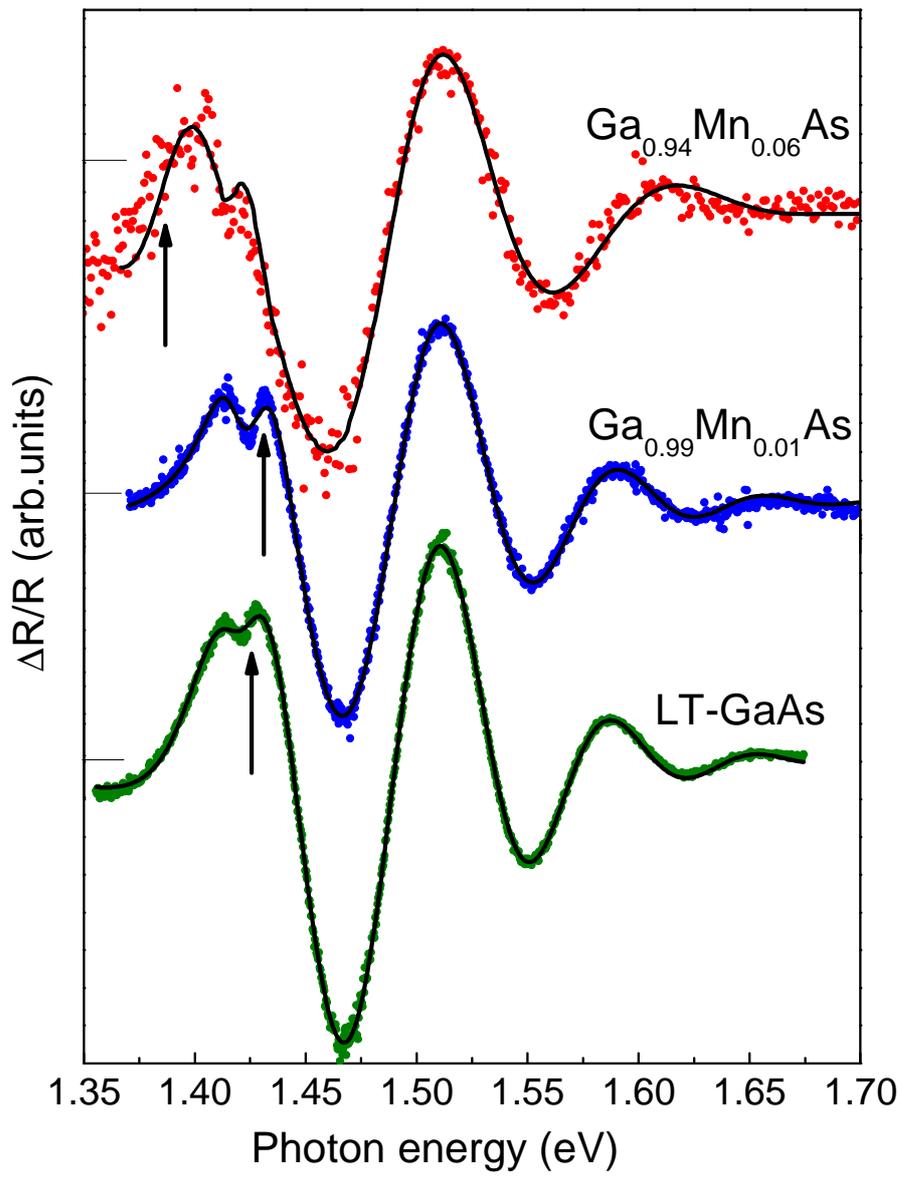

Fig. 6